\documentclass[aps, 
prd,
twocolumn,
superscriptaddress,
showpacs,preprintnumbers,amsmath,amssymb,
longbibliography,floatfix]{revtex4-2}
\usepackage[english]{babel}
\usepackage[utf8]{inputenc}
\usepackage[colorinlistoftodos, color=green!40, prependcaption]{todonotes}
\usepackage{amsthm, bm}
\usepackage{mathtools}
\usepackage{physics}
\usepackage{xcolor}
\usepackage{graphicx}
\usepackage[left=18mm,right=18mm,top=35mm,columnsep=15pt]{geometry} 
\usepackage{adjustbox}
\usepackage{placeins}
\usepackage[T1]{fontenc}
\usepackage{lipsum}
\usepackage{csquotes}
\usepackage[
colorlinks=true, 
pdfstartview=FitV, 
linkcolor=blue, 
citecolor=magenta, 
urlcolor=blue, 
bookmarks=true,
bookmarksnumbered=true,
pdftex,
pdftitle={Article},
pdfauthor={Author}
]{hyperref}
\usepackage[normalem]{ulem}  % \sout{old text} for strikeout
\usepackage{comment}

\usepackage{tikz}
\usepackage[compat=1.1.0]{tikz-feynhand}

\newcommand{\comYA}[1]{{\color{red}[YA:\,#1]}}

\newcommand{\comMS}[1]{{\color{teal}[MS:\,#1]}}

\newcommand{\cout}[1]{ \if 0 {#1} \fi }

\begin{document}
\title{
Enhancement of photon emission rate near QCD critical point
}

\author{Yukinao Akamatsu}
    \email[Correspondence email address: ]{yukinao.a.phys@gmail.com}
\author{Masayuki Asakawa}
    \email{yuki@phys.sci.osaka-u.ac.jp}
    \affiliation{Department of Physics, Osaka University, Toyonaka, Osaka 560-0043, Japan}
\author{Masaru Hongo}
    \email{hongo@phys.sc.niigata-u.ac.jp}
    \affiliation{Department of Physics, Niigata University, Niigata 950-2181, Japan}
    \affiliation{
    RIKEN Center for Interdisciplinary Theoretical and Mathematical Sciences (iTHEMS), RIKEN, Wako 351-0198, Japan}
\author{Mikhail Stephanov}
    \email{misha@uic.edu}
    \affiliation{Department of Physics, University of Illinois, Chicago, IL 60607, USA}
    \affiliation{Laboratory for Quantum Theory at the Extremes, University of Illinois, Chicago, Illinois 60607, USA}
\author{Ho-Ung Yee}
    \email{hyee@uic.edu}
    \affiliation{Department of Physics, University of Illinois, Chicago, IL 60607, USA}
\affiliation{Laboratory for Quantum Theory at the Extremes, University of Illinois, Chicago, Illinois 60607, USA}
\date{\today}

\begin{abstract}
We compute photon emission rate enhancement near the QCD critical point using an effective theory of dynamic critical phenomena and derive a universal photon spectrum.
The emission rate scales similarly to conductivity, increasing with the correlation length ($\xi$), diverging at the critical point. 
The spectrum exhibits $\omega dN_{\gamma}/d^3k \propto \omega^{-1/2}$ in the scaling regime, with the transition occurring at a frequency comparable to shear damping rate $\omega \sim \gamma_{\eta}/\xi^2$, reflecting the nonequilibrium properties of the near-critical liquid.
\end{abstract}


\maketitle

\section{Introduction}
%{\it Introduction ---}
In thermodynamics, a critical point represents the endpoint of a first-order phase transition and plays a crucial role in characterizing the phase structure of physical systems.
The properties of these points adhere to universal laws, determined by only a few microscopic theory characteristics such as symmetries and conservation laws. 
In the study of matter governed by strong interactions, as described by Quantum Chromodynamics (QCD), several critical points have been proposed~\cite{Fukushima:2010bq}.
Among these, the point predicted in the region of temperatures and densities of order the typical strong interaction scale $\mathcal O(100)$ MeV draws the most attention~\cite{Asakawa:1989bq, Stephanov:2004wx} due to the possibility of its discovery in heavy-ion collisions \cite{Stephanov:1998dy, Bzdak:2019pkr}.
The search for this QCD critical point is actively pursued through the Beam-Energy Scan programs at the Relativistic Heavy-Ion Collider (RHIC) \cite{STAR:2010vob} and is planned for future experimental facilities~\cite{Bluhm:2020mpc}.

Promising signatures of the critical point in relativistic heavy-ion collisions based on the non-monotonic behavior of fluctuations were proposed ~\cite{Stephanov:1998dy, Stephanov:1999zu}. These signatures come with challenges and limitations ~\cite{Asakawa:2019kek}. 
Among those, the most obvious one is due to the hydrodynamic expansion inherent in heavy-ion collisions, which hinders the system from achieving full criticality ~\cite{Berdnikov:1999ph}.
One of the strategies to overcome this challenge is to analyze higher-order (3rd and 4th order) cumulants of the baryon number, which are expected to be more sensitive to the critical point effects~\cite{Stephanov:2008qz}. It is clear that a dynamic modeling of heavy-ion collisions, including the effects of criticality on fluctuations is needed.
Various research groups have been developing corresponding analytical and numerical approaches~\cite{Berges:2009jz,Schlichting:2019tbr,Schweitzer:2020noq,Schweitzer:2021iqk,Stephanov:2017ghc,Nahrgang:2018afz,An:2019csj,An:2020vri,Schaefer:2022bfm,Chattopadhyay:2023jfm,Chattopadhyay:2024jlh,Chattopadhyay:2024bcv,Florio:2021jlx,Florio:2023kmy,Basar:2024qxd,Bhambure:2024axa,Bhambure:2024gnf,Stephanov:2024mdj}.
One might anticipate the emergence of a universal nonequilibrium scaling regime in the slow expansion limit offering a potential framework for studying dynamic critical phenomena
~\cite{Chandran:2012cjk,Mukherjee:2015swa,Mukherjee:2016kyu,Akamatsu:2018vjr}.

Complementary information could be gained through the electromagnetic probes, which experience practically no interaction with strongly interacting matter and thus retain information from the production point.
Recent {\em model} calculations have indicated a potential divergence in the production rate of dileptons with low energy and momentum.
It is attributed to the critical behavior of electric conductivity~\cite{Nishimura:2023oqn, Nishimura:2024kvz}, which, in these models, is different from that of model H expected to describe universal critical dynamics of the QCD critical point~\cite{Son:2004iv}. 
The difference comes from the lack of corresponding effective infrared dynamics in terms of soft modes, obscuring the physical interpretation of such a divergence.

\begin{figure}
\includegraphics[width=0.48\textwidth]{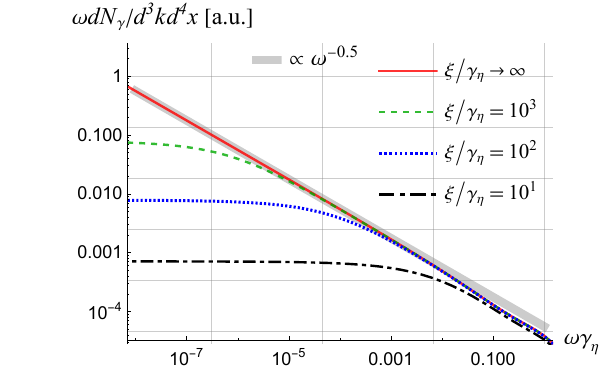}
\caption{Photon spectra from critical fluctuations near the critical point for various correlation length $\xi/\gamma_\eta$.
The absolute magnitude is proportional to an unknown coefficient; see the later discussion around Eq.~\eqref{eq:rate}.
We assume that bare transport coefficients satisfy $\bar\lambda\equiv K\lambda/\gamma_{\eta}^3=1$. 
}
\label{fig:photon}
\end{figure}

In this paper, we predict the photon emission rate from matter near the QCD critical point without relying on any microscopic models (Fig.~\ref{fig:photon}).
Instead, we utilize the universality of the critical fluid, as described by a framework known as model H in the Hohenberg-Halperin classification~\cite{RevModPhys.49.435,Son:2004iv}, to calculate the spectrum.
We find that it diverges as $\omega dN_{\gamma}/d^3k \propto \omega^{-1/2}$ at low energies $\omega = k$.
Slightly away from the critical point this divergence is regularized to $\omega dN_{\gamma}/d^3k \propto \xi$ at $\omega \sim \gamma_{\eta}/\xi^2$.
Here $\gamma_{\eta}=\bar\eta/w$ represents the diffusion coefficient of the transverse momentum ($\bar\eta$ is the shear viscosity and $w=e+p$ is the enthalpy density) and $\xi$ denotes the correlation length, which diverges at the critical point.
This transition occurs when the photon frequency $\omega$ matches the shear damping rate of the critical fluid, $\gamma_{\eta}/\xi^2$, indicating a universal expression in terms of $\nu\equiv\omega\xi^2/\gamma_{\eta}$ for $\omega\ll T$:
\begin{align}
\label{eq:universal}
\omega \frac{dN_{\gamma}}{d^3kd^4x}
\propto\frac{T^2\xi}{\gamma_{\eta}}
\cdot\frac{1-(1-\nu)\sqrt{\nu/2}}{1+\nu^2}
=\frac{T^2\xi}{\gamma_{\eta}}\Phi(\nu)\,.%, \quad
%\Phi(\nu)=\frac{1-(1-\nu)\sqrt{\nu/2}}{1+\nu^2}.
\end{align}
Here, the universal scaling function $\Phi(\nu)$ incorporates frequency and transport coefficients, reflecting the nonequilibrium properties of the critical fluid probed by the on-shell photon, $\omega = k$.
This stands in stark contrast to the critical damping rate observed in quasielastic photon scattering experiments~\cite{PhysRevA.8.2586}, which probe the near-static regime $\omega \simeq 0$ with fixed $\bm k$ and the longitudinal channel instead of the transverse channel of velocity fluctuations.
These experiments are beautifully described by the universal function $\Omega_K(y)=\frac{3}{4}y^{-2}[1+y^2+(y^3-y^{-1})\arctan y]$ with $y=k\xi$, known as the Kawasaki function~\cite{KAWASAKI19701}.
For comparison with our analysis, a derivation of the Kawasaki function is presented in Appendix \ref{app:kawasaki}.

This paper is organized as follows.
In Section \ref{sec:critical_dynamics}, the model H, which describes the critical dynamics of the QCD critical point, is reviewed.
In Section \ref{sec:photon_rate}, the photon emission rate is computed at one-loop level of thermal fluctuations.
Since the photon rate is defined not at a static momentum $k^{\mu}=(0,\bm k)$ but at an on-shell momentum $(|\bm k|,\bm k)$, the effect of sound mode is discussed, which is neglected in the model H due to its fast oscillation.
In Section \ref{sec:discussion}, we summarize the paper.
 
\section{Critical Dynamics}\label{sec:critical_dynamics}
%{\it Critical Dynamics ---}
Hydrodynamics near the QCD critical point is governed by conserved densities and the order parameter.
Due to the mixing of the conserved densities and order parameter fields, it suffices to consider the conserved densities in the description at long time and length scales \cite{Son:2004iv}.
Specifically, these are the density fluctuations of baryon number, energy, and momentum.
Among these five modes, the slowest mode is the specific entropy $\hat s\equiv s/n$ -- a diffusive scalar mode whose fluctuations can be represented by a linear combination of the energy ($\delta e$) and baryon  ($\delta n$) density fluctuations $\delta\hat s\equiv\delta(s/n) = \frac{1}{nT}\delta e - \frac{e+p}{n^2T}\delta n$. The evolution of $\delta\hat s$ is delayed by critical slowing down.
The other modes, such as sound and shear modes, evolve much faster than the diffusive mode at long wavelengths.
At the nonlinear level, the relevant interaction for the diffusive mode $\delta\hat s$, denoted as $\psi$ in the model H, arises from the mode coupling with the shear mode $\bm v_T$, which is the transverse projection of $\bm v$.
This leads us to critical hydrodynamics, also known as model H~\cite{Son:2004iv}.

The Langevin equation of model H is written as~\cite{RevModPhys.49.435}
\begin{subequations}\label{eq:H}
\begin{align}
\dot{\psi } 
&= -\frac{1}{w}\bm\nabla\cdot\left(
\psi \frac{\delta F}{\delta \bm v_T}
\right) + \lambda\nabla^2\frac{\delta F}{\delta\psi} + \bm\nabla\cdot\bm \zeta_{\psi}, \\
%&= -\vec\nabla\cdot(\psi \vec v_T) + \bar\lambda\nabla^2\frac{\delta F}{\delta\psi} + \vec\nabla\cdot\vec \zeta_{\psi}, \\
\label{eq:pi-dot}
w\dot{\bm v}_{T} &=P_T\left(
-\psi\bm\nabla\frac{\delta F}{\delta\psi}
+\frac{\bar\eta}{w}\nabla^2 \frac{\delta F}{\delta\bm v_T} + \bm\zeta_{v}\right),\\
 F  &= \int \hspace{-3pt} d^3x 
 K\bigg[
  \frac{(\bm\nabla\psi)^2}{2} + \frac{a\psi^2}{2}
    +\frac{b_3\psi^3}{3} + \frac{b_4\psi^4}{4} 
 \bigg] + \frac{w v_T^2}{2} \hspace{-1pt}.
\end{align}
\end{subequations}
Here, $K, a, b_3, b_4$ are parameters of the effective potential, $\lambda=w^2\sigma_B/(n^4T^2)$ is proportional to baryon conductivity $\sigma_B$, and $P_T (\bm V)$ denotes the transverse projection of a vector $\bm V$, i.e. $P_T(\bm V)\equiv \bm V - (\bm V\cdot\bm k)\bm k/\bm k^2$ in  $\bm k$-space.
The property of noises $\bm\zeta_{\psi}$ and $\bm\zeta_{v}$ is determined by requiring the fluctuation-dissipation relation:
\begin{subequations}
\begin{align}
\label{eq:noise_psi}
\langle\zeta_{\psi k}(x)\zeta_{\psi l}(x')\rangle &=2T\lambda\delta_{kl}\delta^4(x-x'),\\
\langle\zeta_{v k}(x)\zeta_{v l}(x')\rangle &= -2T\bar\eta \delta_{kl}\nabla^2\delta^4(x-x').
\end{align}
\end{subequations}
The parameters satisfy $K=n^2T/(c_p a)$ and $a=1/\xi^2$, where $c_p$ is isobaric specific heat \footnote{
This relation is derived by comparing the free energy density at quadratic order of $\psi$
\begin{equation*}
    f= \frac{K}{2}\left[(\bm\nabla \psi)^2 + a\psi^2\right]
     =\frac{n^2T}{2c_p}\left[\xi^2(\bm\nabla \psi)^2 + \psi^2\right].
\end{equation*}
}.
If we neglect the self-interaction of $\psi$, $c_p\propto \xi^2$ and thus $K$ approaches a constant at the critical point.
The self-interactions modifies the relation to $c_p\propto \xi^{2-\eta} = a^{-1+\eta/2}$, where the scaling exponent is $\eta=\mathcal O(\epsilon^2)$ in the $\epsilon$-expansion scheme ($\epsilon=4-d$ and $d$ is the spatial dimension). 
A combination $K\lambda a$ is the (heat) diffusion coefficient, expressed using baryon conductivity by $D=K\lambda a = w^2 \sigma_B / (Tn^2 c_p)$.

\section{Photon Emission Rate}\label{sec:photon_rate}
%{\it Photon Emission Rate ---}
Formula for photon emission rate from medium with temperature $T$ is~\cite{LeBellac_1996}
\begin{align}
k\frac{dN_{\gamma}}{d^3k d^4x} = \frac{\alpha}{2\pi^2}\frac{{\rm Im}\Pi^{T ii}_{R}(k,\bm k)}{e^{k/T}-1}
\simeq \frac{\alpha}{2\pi^2}\Pi^{T ii}_{S}(k,\bm k), 
\end{align}
where $\Pi_{R/S}^{T ii}(\omega,\bm k)=(\delta_{ij}-k_ik_j/\bm k^2)\Pi^{ij}_{R/S}(\omega,\bm k)$ is trace of transverse projection of the response/correlation function $\Pi_{R/S}^{ij}(\omega,\bm k)$.
When $\omega\ll T$, the fluctuation-dissipation theorem dictates $\Pi_{S}^{Tij}(\omega, \bm k) \simeq \frac{T}{\omega}{\rm Im}\Pi_{R}^{Tij}(\omega, \bm k)$.
The correlation function is defined by the electric current $\bm J_{\rm em}$ (in unit of the elementary charge $e = \sqrt{4\pi\alpha}$)
\begin{align}
\Pi_S^{ij}(\omega,\bm k) \equiv \int d^4x e^{i\omega t - i\bm k\cdot\bm x}
\langle J_{\rm em}^i(x)J_{\rm em}^j(0)\rangle_{T,\mu} .
\end{align}
Corresponding to the transverse polarization of photons, only the transverse components contribute to the emission rate.
Here, our focus lies on the soft momentum region, which is predominantly influenced by critical fluctuations.
The electric current in $N_f=2$ is decomposed as $J_{\rm em}^{\mu} = \frac{1}{2}J_{\rm B}^{\mu} + J_{\rm I}^{\mu}$, where $J_{\rm B}^{\mu}$ represents the baryon current and $J_{\rm I}^{\mu}$ denotes the isospin current. 
Furthermore, near the critical point, the isospin fluctuation is shown to be irrelevant~\cite{Son:2004iv}, leading to the approximation $J_{\rm em}^{\mu} \simeq \frac{1}{2}J_{\rm B}^{\mu}$.
Hence, it becomes necessary to define the baryon current $J_{\rm B}^{\mu}$ within the framework of model H.

In hydrodynamics, baryon current is expanded by fluctuations and their derivatives
\begin{subequations}
\label{eq:current}
\begin{align}
\label{eq:current_0}
J_B^0 &= n  
+ \left(\frac{\partial n}{\partial\hat s}\right)_{p}\delta\hat s 
+ \left(\frac{\partial n}{\partial p}\right)_{\hat s}\delta p
+ \frac{n v^2}{2} + \cdots, \\
\label{eq:current_i}
\bm J_B &= \left[n + \left(\frac{\partial n}{\partial\hat s}\right)_{p}\delta\hat s
+ \left(\frac{\partial n}{\partial p}\right)_{\hat s}\delta p
\right]\bm v %\nonumber \\ 
-\sigma_B T \bm\nabla \left(\frac{\mu}{T}\right) + \cdots,
\end{align}
\end{subequations}
where the thermodynamic derivatives can be expressed by $(\partial n/\partial\hat s)_p = -n^2(1-\gamma^{-1})/b$ and $(\partial n/\partial p)_{\hat s}=n/(c_s^2 w)$.
Here $b\equiv (\partial p/\partial T)_n$ is finite and the adiabatic index $\gamma\equiv c_p/c_v$ is divergent at the critical point.
The diffusive mode is mostly composed of $\delta \hat s$ while the sound mode mostly consists of the fluctuations of pressure $\delta p$ and longitudinal momentum $\bm v_L$.
In the model H for critical dynamics, the latter decouples due to its fast oscillation.
Furthermore, the squared amplitudes of the transverse momentum $v_T^2$ is negligible compared to critical fluctuation $\delta\hat s$.
Thus, we can write
\begin{subequations}
\begin{align}
J_B^0 &\simeq n + \mathcal A\psi,\\
\bm J_B &\simeq n\bm v_T
+\mathcal A
\left[\psi \bm v_T + K\lambda\bm\nabla\left[(\nabla^2 -a)\psi\right] -\bm \zeta_{\psi} \right],
\end{align}
\end{subequations}
with $\mathcal A\equiv (\partial n/\partial\hat s)_p$.
\cout{
\comYA{Previous definition: $\mathcal A =  \left(\frac{\partial n}{\partial\hat s}\right)_p\frac{1}{\sqrt{K}}$.}
}

\subsection{One-loop Calculation of Model H}
%{\it One-loop Calculation of Model H ---}
Using the Langevin equation of model H, we calculate the correlation function of the transverse electric current $\Pi_{S}^{T ij}(\omega, \bm k)$ at the one-loop level.
Since the transverse electric current is given by
\begin{align}
\bm J_{\rm em}^{\, T}& \equiv P_T(\bm J_{\rm em})
\simeq \frac{\mathcal A}{2}  P_T\left(\psi \bm v_T -\bm \zeta_{\psi}\right) + \frac{n}{2} \bm v_T,
\end{align}
the two-point function consists of various combinations of terms.
At tree level, $\langle \zeta_{\psi k}(x)\zeta_{\psi l}(x')\rangle$ and $\langle v_{Tk}(x) v_{Tl}(x')\rangle$ contribute and we obtain for light-like momenta $k^{\mu} $ as $\Pi_{S}^{T ii}(k, \bm k)|_{\rm tree} \simeq {\mathcal A}^2 T\lambda + \frac{n^2}{w}T\gamma_{\eta}$ in the limit $\gamma_{\eta}k\ll 1$.
At the one-loop level, we need to calculate $\langle \psi v_{Tk}(x) \cdot\psi v_{Tl}(x')\rangle$, one-loop correction to $\langle v_{Tk}(x) v_{Tl}(x')\rangle$, and $\langle\psi v_{Tk}(x)\cdot\zeta_{\psi l}(x')\rangle$, while the other terms contribute only at the two-loop level and beyond.
For light-like momenta $k^{\mu} $, we obtain for $\langle \psi v_{Tk} \cdot\psi v_{Tl}\rangle$
\begin{align}
\label{eq:PiT_oneloop}
&\Pi_{S}^{T ii}(k, \bm k)|_{\text{1-loop}} \simeq 
\scalebox{0.8}{
  \begin{tikzpicture}[baseline=(o.base)]
   \begin{feynhand}
    \vertex (o) at (0,-0.1) {};
    \vertex (a0) at (-1.6,0) {};
    \vertex (b0) at (1.6,0) {};
    \vertex [dot] (a) at (-0.8,0) {};
    \vertex [dot] (b) at (0.8,0) {};
    \vertex at (-1.2,0.5) {$k$};
    \vertex at (1.2,0.5) {$k$};
    \vertex at (-1,-0.3) {$i$};
    \vertex at (1,-0.3) {$j$};
    \propag [mom'={[arrow shorten=0.33] $p$}] (a) to [out=90,in=90, looseness=1.3] (b);
    \propag [sca,mom={[arrow shorten=0.33] $k-p$}] (a) to [out=270,in=270, looseness=1.3] (b);
    \propag [photon,mom={[arrow shorten=0.33]},gray] (a0) to (a);
    \propag [photon,mom={[arrow shorten=0.33]},gray] (b) to (b0);
   \end{feynhand}
  \end{tikzpicture}}
\nonumber \\
&\quad = \frac{{\mathcal A}^2}{4} \left(\delta_{ij}-\frac{k_ik_j}{\bm k^2}\right)
\int\frac{d^4p}{(2\pi)^4} G_{S0}^{\psi\psi}(p) G_{S0}^{T ij}(k-p), 
\end{align}
where $G^{\psi\psi}_{S0}$ and $G^{T ij}_{S0}$ are tree-level correlation functions given by
\begin{subequations}
\label{eq:GG}
\begin{align}
G_{S0}^{\psi\psi}(p) &=
  \scalebox{0.8}{
  \begin{tikzpicture}[baseline=(o.base)]
   \begin{feynhand}
    \vertex (o) at (0,-0.1) {};
    \vertex [dot] (a) at (-0.6,0) {};
    \vertex [dot] (b) at (0.6,0) {};
    \propag [mom={[arrow shorten=0.25] $p$}] (a) to (b);
   \end{feynhand}
  \end{tikzpicture}}
= \frac{2T\lambda \bm p^2}{(p^0)^2 + (K\lambda)^2 \bm p^4(a + \bm p^2)^2},\\
G_{S0}^{T ij}(p) &=
  \scalebox{0.8}{
  \begin{tikzpicture}[baseline=(o.base)]
   \begin{feynhand}
    \vertex (o) at (0,-0.1) {};
    \vertex [dot] (a) at (-0.6,0) {};
    \vertex [dot] (b) at (0.6,0) {};
    \vertex at (-0.6,0.3) {$i$};
    \vertex at (0.6,0.3) {$j$};
    \propag [sca, mom={[arrow shorten=0.25] $p$}] (a) to (b);
   \end{feynhand}
  \end{tikzpicture}}
 =
\frac{2Tw^{-1}\gamma_{\eta}(\delta_{ij}\bm p^2 - p_ip_j)}{(p^0)^2 + \gamma_{\eta}^2 \bm p^4}. 
\end{align}
\end{subequations}
As shown below, Eq.~\eqref{eq:PiT_oneloop} behaves as $\propto \omega^{-1/2}$ in the low photon energy $\omega=k\to 0$ at the critical point $a=0$, contributing to the renormalization of $\lambda$.
The next singular contribution comes from one-loop correction to $\langle v_{Tk}(x)v_{Tl}(x')\rangle$, which behaves as $\propto \omega^{-1/4}$ and contributes to the renormalization of $\gamma_{\eta}$, which will be reported in a separate paper.
Finally, $\langle\psi v_{Tk}(x)\cdot\zeta_{\psi l}(x')\rangle$ does not yield any singular contribution at low energy.

When both $a=1/\xi^2$ and $\omega$ approach 0 near the critical point, a natural scaling variable is the ratio between the photon frequency $\omega$ and the typical damping rate $\gamma_{\eta}/\xi^2$ of $\bm v_T$, yielding $\nu=\omega\xi^2/\gamma_{\eta}=\omega /(\gamma_{\eta}a)$.
Note that the damping rate of $\bm v_T$ with the wavelength $\xi$ is relevant because of the mode coupling between $\psi$ and $\bm v_T$.
The one-loop integral near the critical point can be approximated as
\begin{align}
\label{eq:PiT_oneloop_result}
\Pi_{S}^{T ii}(k, \bm k)|_{\text{1-loop}}
&\simeq \frac{\mathcal A^2 T^2}{6\pi K w\gamma_{\eta}a^{1/2}}\Phi(\nu).
\end{align}
The left hand side can be expressed using a dimensionless function $F(\nu,\bar a,\bar \lambda)$ with three scaled variables $\nu$, $\bar a = a\gamma_{\eta}^2$, and $\bar\lambda=K\lambda/\gamma_{\eta}^3$:
\begin{align}
    \Pi_{S}^{T ii}(k, \bm k)|_{\text{1-loop}}
    =\frac{\mathcal A^2}{4}\frac{T^2}{2\pi^2 K w}
    F(\nu, \bar a, \bar\lambda).
\end{align}
Thus the critical scaling behavior \eqref{eq:PiT_oneloop_result} is equivalent to
\begin{align}
    \lim_{\bar a\to 0}\frac{3\bar a^{1/2}}{4\pi}F(\nu,\bar a,\bar\lambda)
    =\Phi(\nu).
\end{align}
Near the critical point, $F(\nu,\bar a,\bar\lambda)$ is independent of $\bar\lambda$ because the soft mode $\psi$ is effectively frozen due to the critical slowing down.

\begin{figure}
\includegraphics[width=0.45\textwidth]{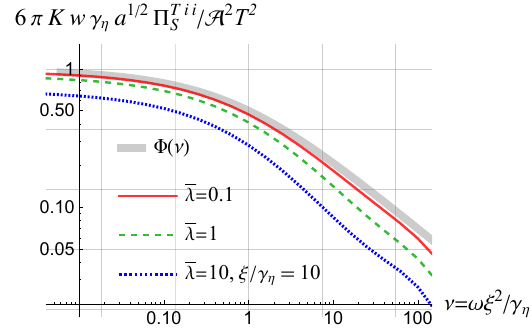}
\caption{$\Phi(\nu)$ is compared with Eq.~\eqref{eq:PiT_oneloop} calculated using Eq.~\eqref{eq:GG} with $\bar a\equiv a\gamma_{\eta}^2=0.01$ ($\xi/\gamma_{\eta}=10$) and $\bar\lambda\equiv K\lambda/\gamma_{\eta}^3 = 0.1, 1, 10$.
This corresponds to comparing $\Phi(\nu)$ and $\frac{3\bar a^{1/2}}{4\pi}F(\nu,\bar a,\bar\lambda)$.
}
\label{fig:scaling}
\end{figure}

In Fig.~\ref{fig:scaling}, $\Phi(\nu)$ is compared with the numerical integration of the left hand side of \eqref{eq:PiT_oneloop_result}, with a proper multiplicative constant.
A weak dependence on $\bar\lambda$ is found for a fixed $\bar a=0.01$.
By using the asymptotic forms of $\Phi(0)=1$ and $\Phi(\nu\gg 1)\simeq 1/\sqrt{2\nu}$, we obtain a formula for the critical enhancement of the photon emission rate for $\omega = k \ll T$ (see Appendix \ref{app:oneloop_detail} for details):
\begin{align}
\label{eq:criticalphoton}
\hspace{-2mm}
\omega\frac{dN_{\gamma}}{d^3kd^4x} \simeq \left\{\begin{aligned}
& \frac{\alpha{\mathcal A^2}}{12\pi^3 K w}\frac{T^2\xi}{\gamma_{\eta}} & \left(\omega \ll \frac{\gamma_{\eta}}{\xi^2}\right) \\
& \frac{\alpha{\mathcal A^2}}{12\pi^3 K w}\frac{T^2}{\sqrt{2\gamma_{\eta} \omega}} & 
\left(\frac{\gamma_{\eta}}{\xi^2} \ll \omega \ll \frac{c_s^2}{\gamma_{\eta}}\right)
\end{aligned}\right.
\end{align}
Around $\omega\sim c_s^2/\gamma_{\eta}$, we need to consider the sound mode contributions to the photon production as will be discussed below.

The emission rate for soft photons increases as $\omega^{-1/2}$ in the infrared until it reaches $\xi/\gamma_{\eta}$ at $\omega\sim \gamma_{\eta}/\xi^2$. 
In the presence of photon background, the system is driven out of equilibrium for larger frequencies, while it remains in the Ohmic regime for smaller frequencies.
The transition occurs when the frequency balances the typical damping rate of $\bm v_T$, i.e. $\omega=\gamma_{\eta}/\xi^2$.
Thus, the photon spectra reflect the nonequilibrium dynamical properties of the critical liquid. 
In the hydrodynamic regime, as $\omega\to0$, the correlation function behaves $\Pi_S^{Tii}\simeq\mathcal A^2T[\lambda + T\xi/(6\pi K \bar\eta)]=\mathcal A^2T\lambda_{\rm hydro}$, leading to $\lambda_{\rm hydro}\propto\xi$.
Thus, the diffusion coefficient $D=K\lambda/\xi^2$ is renormalized as $D_{\rm hydro}=D+T/(6\pi\bar\eta\xi)\propto 1/\xi$.
This relationship is well-known and leads to the determination of the dynamic critical exponent of the model H as approximately $z\simeq 3$~\cite{PhysRevB.13.2110, 10.1143/PTP.55.1384}.

\subsection{Sound Mode Contribution}
%{\it Sound Mode Contribution ---}
Here we evaluate the contribution of sound mode to the photon production rate.
In the model H, the sound mode decouples from the slow diffusive dynamics because of its fast oscillatory phase with frequency $k^0=c_s|\bm k| \gg \gamma_{\eta}\bm k^2$.
However, when we consider on-shell photon production, photon involves an even larger frequency $k^0 = |\bm k|$ and thus the sound mode does not necessarily decouple.

The baryon current \eqref{eq:current_i} contains a nonlinear coupling $\propto\delta\hat s \bm v$ as well as a linear term $\propto \bm v$, which both include the sound mode $\bm v_L$ (longitudinal projection of $\bm v$).
Although there is no tree contribution from the sound mode to the photon production rate due to the transverse projection, the sound mode contributes from the one-loop order.
The nonlinear coupling between the sound mode $\bm v_L$ and the critical mode $\delta\hat s$ may enhance the photon production rate via the thermal loop effect.

The one-loop integral is similar to Eq.~\eqref{eq:PiT_oneloop}.
One only needs to replace the transverse correlator $G^{Tij}_{S0}(p)$ with the longitudinal one:
\begin{align}
G^{Lij}_{S0}(p)
=\frac{2Tw^{-1}\gamma_\zeta (p^0)^2 p_ip_j}{[(p^0)^2-c_s^2\bm p^2]^2+(\gamma_\zeta p^0\bm p^2)^2}.
\end{align}
For photon frequencies $\omega \lesssim c_s^2/\gamma_{\eta}$, the one-loop integral is found to be $\propto a^0$, showing no critical enhancement in the photon emission rate $\omega dN_{\gamma}/d^3k\propto 1/c_s$ (see Appendix \ref{app:sound_detail} for details).

\begin{figure}
\includegraphics[width=0.5\textwidth]{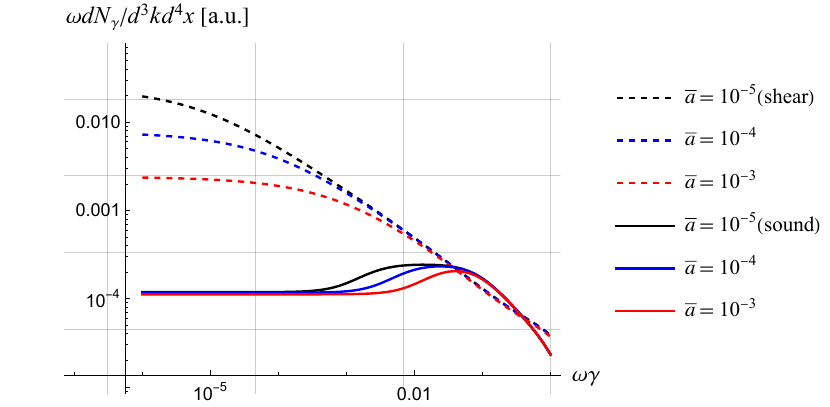}
\caption{Photon spectrum from sound mode loop is compared with that from model H calculation near the critical point $\bar a\equiv a\gamma_{\eta}^2=10^{-5}, 10^{-4}, 10^{-3}$.
The absolute magnitude is proportional to an unknown coefficient; see the later discussion around Eq.~\eqref{eq:rate}.
%The spectra are multiplied by an unknown coefficient $Kw/\mathcal A^2 T^2 \simeq w \xi^2/\chi_T T^2$.
We assume that bare transport coefficients satisfy $\bar\lambda\equiv K\lambda/\gamma_{\eta}^3=1$ and $\gamma_{\eta}=\gamma_{\zeta}/2$, and that sound velocity is $c_s^2 = 1/3$.
%The estimate in \eqref{eq:sound_estimate} corresponds to the photon yield $\sim 10^{-5}$ at $\omega=0$, which is about 10\% of the numerical integration. 
}
\label{fig:photon_sound}
\end{figure}

The photon emission rate from sound mode is numerically calculated in Fig.~\ref{fig:photon_sound} and compared with the model H results.
%We assume the transport coefficients satisfy $\bar\lambda\equiv K\lambda/\gamma_{\eta}^3=1$ and $\gamma_{\eta} = \gamma_{\zeta}/2$, and the sound velocity is $c_s^2 = 1/3$.
The overall magnitude from the sound loop is $\propto a^0$, showing no enhancement with $a\to 0$.
In contrast, the model H calculation shows critical enhancement with $\omega dN_{\gamma}/d^3k\propto (\gamma_{\eta}\omega)^{-1/2}$ for $\omega \gg \gamma_{\eta}/\xi^2$ and dominates over the sound mode contributoin for $\omega\gamma_{\eta} \ll c_s^2$.
Therefore, the scaling behavior of model H calculation persists in a regime $\gamma_{\eta}/\xi^2 \ll \omega \ll c_s^2/\gamma_{\eta}$.

\section{Discussion and Outlook}\label{sec:discussion}
%{\it Discussion and Outlook ---}
In this paper, we have derived a formula \eqref{eq:criticalphoton} for the enhanced emission rate of soft photons from a critical liquid. 
It is expressed using a universal function $\Phi(\nu)$ [see Eq.~\eqref{eq:universal}], which characterizes the nonequilibrium property of the critical liquid probed by on-shell photons.
Around $\omega\sim 0$, we estimate the magnitude of the critical photon spectra from a spacetime volume $V_4$ by:
\begin{align}
    &\omega\frac{dN_{\gamma}}{d^3k}\Biggl|_{\rm crit} \simeq \frac{\alpha}{12\pi^3} \cdot \frac{T^2}{\bar\eta}\cdot \frac{\chi_T}{\xi^2}\cdot \xi V_4.
 \label{eq:rate}
\end{align}
Here we use the relations near the critical point, $\mathcal A^2/K = n^2 (1-\gamma^{-1})^2 c_p a/ (T b^2) \simeq n^2 c_p/(T b^2\xi^2)$ and $T\chi_T = (1-\gamma^{-1})n^2 c_p/b^2\simeq n^2 c_p/b^2$, where $\chi_T\equiv (\partial n/\partial \mu)_T$ is the baryon number susceptibility.
Note that $\chi_T/\xi^2$ approaches a constant near the critical point when we neglect the self-interaction of $\psi$.
If we include the full effects of critical fluctuations, the scaling would be estimated by $\xi^{1-\eta-x_{\eta}}=\xi^{x_{\lambda}}$.
Here $x_{\lambda}$ and $x_{\eta}$ are scaling exponents of $\lambda$ and $\bar\eta$ and are calculated as $x_{\lambda} = 18\epsilon/19 + \mathcal O(\epsilon^2)$ and $x_{\eta} = \epsilon/19 + \mathcal O(\epsilon^2)$ in the $\epsilon$-expansion scheme.
There is an uncertainty in the overall magnitude due to the unknown constant $\chi_T/\xi^2$ (or $\chi_T/\xi^{2-\eta}$), which makes it difficult to predict the absolute magnitude.
A crude estimate would be made by dimensional analysis.
Since the ratio $\chi/\xi^2$ has mass dimension 4, typical scale is given by QCD scale such as the temperature $T_c$.
Then $\omega dN_{\gamma}/d^3k\sim \alpha (T^2/\bar\eta) T^4 \xi V_4 \sim \alpha T^3\xi V_4$ for $\bar \eta\sim T^3$.
As a reference, the photon from normal QGP can also be estimated as $\omega dN_{\gamma}/d^3k\sim\alpha T^2 V_4$.
The critical enhancement is given by a factor $\xi T\sim 3$ for $\xi\sim 3{\rm fm}$ and $T\sim 1{\rm fm^{-1}}$.
As for the scaling scaling $\omega dN_{\gamma}/d^3k \propto \omega^{-1/2}$, it is predicted robustly.
For example, we can distinguish from the photon spectrum due to microscopic scattering in a weakly coupled quark-gluon plasma, which is predicted to be $\omega dN_{\gamma}/d^3k \propto \omega^{-3/2}$ for photons with low frequencies $\omega$, such that $eT, g_s^4 T\ln g_s^{-1}\ll \omega\ll T$~\cite{Arnold:2001ms}, where $g_s$ is the QCD coupling constant.

Although observing the critical enhancement of low-energy photons may be challenging in heavy-ion collisions, our formula with the universal function $\Phi(\nu)$ holds significant theoretical value.
Since photons probe lightlike kinematic regimes, the spectrum reflects the nonequilibrium properties of the critical fluid. 
This is indicated by the variable $\nu = \omega \xi^2/\gamma_{\eta}$ in the universal function $\Phi(\nu)$, which incorporates the photon frequency $\omega$ and the transport coefficient $\gamma_{\eta}$.

The result presented in this paper highlights the most singular behavior of the photon spectrum.
However, there exists another singular contribution arising from the momentum density fluctuation $\langle v_{Tk}(x)v_{Tl}(x')\rangle$, which renormalizes $\gamma_{\eta}$ (or $\bar\eta$) at one-loop level.
Incorporating this contribution and applying a renormalization group treatment to the photon spectrum would enable us to derive small deviations from the exponent $\omega dN_{\gamma}/d^3k\propto\omega^{-1/2}$.
With this analysis, we can complete the one-loop calculation of the current correlation function with lightlike momentum.
Then, the next target after the photon spectrum would naturally be the dilepton spectrum.

\acknowledgements
We thank Teiji Kunihiro and Masakiyo Kitazawa for useful discussions.
Y.A. and M.H. are supported by Japan Society for the Promotion of Science (JSPS) KAKENHI Grant Numbers JP23H01174 and JP23K25870.
M.H. is supported by Japan Society for the Promotion of Science (JSPS) KAKENHI Grant Numbers 22K20369 and 25K07316, and RIKEN iTHEMS.
M.A. is supported in part by Japan Society for the Promotion of Science (JSPS) KAKENHI Grant Number JP23K03386. M.S. and H.-U.Y. are supported by the U.S. Department of Energy, Office of Science, Office of Nuclear Physics, grant No. DEFG0201ER41195.

\appendix
\begin{widetext}
%\pagebreak
%\widetext
%\begin{center}
%\textbf{\large Supplemental Materials:\\
%Enhancement of photon emission rate near QCD critical point}
%\end{center}
%%%%%%%%%% Merge with supplemental materials %%%%%%%%%%
%%%%%%%%%% Prefix a "S" to all equations, figures, tables and reset the counter %%%%%%%%%%
%\setcounter{equation}{0}
%\setcounter{figure}{0}
%\setcounter{table}{0}
%\setcounter{page}{1}
%\newcounter{supplementeqcountar}
%\makeatletter
%\newcommand\suppsect[1]{{\it #1.}---}%Define section
%\renewcommand{\theequation}{S\arabic{equation}}
%\renewcommand{\thefigure}{S\arabic{figure}}
% \renewcommand{\bibnumfmt}[1]{[S#1]}
% \renewcommand{\citenumfont}[1]{S#1}
%%%%%%%%%% Prefix a "S" to all equations, figures, tables and reset the counter %%%%%%%%%%

%\section*{Supplementary Materials}

\section{Kawasaki function}
\label{app:kawasaki}
It is worth noting that the Kawasaki function $\Omega_K(y)=\frac{3}{4}y^{-2}[1+y^2+(y^3-y^{-1})\arctan y]$~\cite{KAWASAKI19701} is derived from an integral akin to that in Eq.~\eqref{eq:loop_detail} when computing the one-loop correction to the conductivity $\lambda$. 
The key distinctions lie in its evaluation at $k^0=0$ and the longitudinal projection $k^ik^j/\bm k^2$, instead of the transverse projection.
Then, the $\bm k$-dependent one-loop correction $\Delta\lambda(\bm k)$ reads
\begin{align}
\Delta \lambda(\bm k)&=\frac{1}{2T}\frac{k_ik_j}{\bm k^2}\int\frac{d^4p}{(2\pi)^4} G_{S0}^{\psi\psi}(p)G_{S0}^{T ij}(k-p)\Bigr|_{k^0=0}\\
&=\frac{T}{4\pi^2 K w} \int_0^{\infty}p^2 dp \int_{-1}^1 dz
p^2 \left(p^2-p^2z^2\right) \frac{1}{p^2(a + p^2)}\frac{1}{k^2 -2kpz + p^2}
\frac{1}{K\lambda p^2(a + p^2) + \gamma_{\eta} (k^2 -2kpz + p^2)}. \nonumber
\label{eq:loop_kawasaki}
\end{align}
Introducing dimensionless variables $x\equiv p/\sqrt{a}$ and $y\equiv k/\sqrt{a}$, it becomes
\begin{align}
\Delta \lambda(\bm k)&=
\frac{T}{4\pi^2 K w\sqrt{a}}\int_0^{\infty} \frac{x^2 dx}{1 + x^2} 
\int_{-1}^1 dz\frac{1}{y^2 -2yxz + x^2}
\frac{x^2-x^2z^2}{K\lambda a x^2(1 + x^2) + \gamma_{\eta} (y^2 -2yxz + x^2)}.
\end{align}
In the limit as $a \to 0$, while keeping $y \sim 1$, the integral is dominated by the region where $x \sim 1$, allowing us to approximate it as
\begin{align}
\Delta \lambda(\bm k)&\simeq
\frac{T}{4\pi^2 K w\gamma_{\eta}\sqrt{a}}
\int_0^{\infty} \frac{x^4 dx}{1 + x^2} \int_{-1}^1 dz
\frac{1-z^2}{ (y^2 -2yxz + x^2)^2} \nonumber\\
&=\frac{T}{8\pi K w\gamma_{\eta}\sqrt{a}}
\frac{1}{y^2(1+y^2)}\left[
1+y^2+(y^3-y^{-1})\arctan y
\right]
=\frac{T}{6\pi K w\gamma_{\eta}\sqrt{a}}\frac{\Omega_K(y)}{1+y^2}.
\end{align}
The damping rate of $\psi$ near the critical point is predominantly governed by the one-loop contribution, which scales as $a^{-1/2}$, and thus is expressed as
\begin{align}
\Gamma_{\psi}(\bm k) = K(\lambda+\Delta\lambda(\bm k))k^2 (k^2 + a)\simeq K\Delta\lambda(\bm k)k^2(k^2+a).
\end{align}
Using the scaling variable $y=k\xi=k/\sqrt{a}$, it is written as
\begin{align}
\Gamma_{\psi}(\bm k) \simeq 
\frac{T}{6\pi \eta\xi^3}y^2\Omega_K(y).
%=\frac{T}{6\pi\eta\xi^3}y^2\Omega_K(y).
\end{align}
With the asymptotic behaviors $\Omega_K(0)=1$ and $\Omega_K(y\gg 1)\simeq 3\pi y/8$, we can infer that $\lim_{a\to 0}\Gamma_{\psi}(\bm k)\propto k^3$ and $z=3$.

\section{Derivation of Eq.~\eqref{eq:PiT_oneloop_result}}
\label{app:oneloop_detail}
Here, we provide detailed calculation of Eq.~\eqref{eq:PiT_oneloop_result} in the main text.
The propagators are defined by the two-point functions in the linearized approximation, denoted as $\langle\cdots\rangle_0$:
\begin{align}
G_{S0}^{\psi\psi}(p) &\equiv \int d^4xe^{-ip\cdot x}\langle\psi(x)\psi(0)\rangle_0
=\frac{2T\lambda \bm p^2}{(p^0)^2 + (K\lambda)^2 \bm p^4(a + \bm p^2)^2},\\
G_{S0}^{T ij}(p) &\equiv \int d^4xe^{-ip\cdot x}\langle v_{T}^i(x) v_{T}^j(0)\rangle_0
=\frac{2Tw^{-1}\gamma_{\eta}(\delta_{ij}\bm p^2 - p_ip_j)}{(p^0)^2 + \gamma_{\eta}^2 \bm p^4}. 
\end{align}
Then, the one-loop integral is calculated as
\begin{align}
&\left(\delta_{ij}-\frac{k_ik_j}{\bm k^2}\right)\left[\int\frac{d^4p}{(2\pi)^4} G_{S0}^{\psi\psi}(p)G_{S0}^{T ij}(k-p)\right] \nonumber \\
&=\left(\delta_{ij} - \frac{k_ik_j}{\bm k^2}\right)\int\frac{d^4p}{(2\pi)^4}
\frac{2T\lambda\bm p^2}{(p^0)^2 + (K\lambda)^2 \bm p^4(a + \bm p^2)^2}
\frac{2Tw^{-1}\gamma_{\eta}\left[\delta_{ij}(\bm k-\bm p)^2 - (k_i-p_i)(k_j-p_j)\right]}
{(k^0-p^0)^2 + \gamma_{\eta}^2 (\bm k-\bm p)^4} \nonumber \\
&=\frac{4T^2\lambda \gamma_{\eta}}{w}\int\frac{d^3 p}{(2\pi)^3}
\bm p^2 \left((\bm k-\bm p)^2 + \frac{(\bm k^2 - \bm k\cdot\bm p)^2}{\bm k^2}\right) \nonumber \\
&\qquad\qquad\times \left[
\frac{1}{2K\lambda \bm p^2(a + \bm p^2)} +\frac{1}{2 \gamma_{\eta} (\bm k-\bm p)^2}
\right]
\frac{1}{(k^0)^2 + [K\lambda \bm p^2(a + \bm p^2) + \gamma_{\eta} (\bm k-\bm p)^2]^2} \nonumber \\
&=\frac{T^2\lambda\gamma_{\eta}}{2\pi^2 w} \int_0^{\infty}p^2 dp \int_{-1}^1 dz
p^2 \left(k^2 - 2kpz + p^2 + (k-pz)^2\right) \nonumber \\
&\qquad\qquad\times \left[
\frac{1}{K\lambda p^2(a + p^2)}+\frac{1}{\gamma_{\eta} (k^2 -2kpz + p^2)}
\right]
\frac{1}{(k^0)^2 + [K\lambda p^2(a + p^2) + \gamma_{\eta} (k^2 -2kpz + p^2)]^2}.
\label{eq:loop_detail}
\end{align}
From here, we use $\omega = k_0 = k$ for light-like external momentum:
\begin{align}
\label{eq:loop_detail_photon}
\text{\eqref{eq:loop_detail}}\Bigr|_{k_0=k=\omega} &= 
\frac{T^2}{2\pi^2 K w} \int_0^{\infty}dp \int_{-1}^1 dz
\left(\frac{\gamma_{\eta}p^2}{a+p^2} + \frac{K\lambda p^4}{\omega^2-2\omega p z + p^2}\right)
\frac{2\omega^2 - 4\omega pz + p^2 + p^2z^2}{\omega^2 + \left[
K\lambda p^2(a+p^2) + \gamma_{\eta} (\omega^2-2\omega pz + p^2)
\right]^2}\\
&=\frac{T^2}{2\pi^2 K w} F(\omega/(\gamma_\eta a), a\gamma_{\eta}^2, K\lambda/\gamma_{\eta}^3),
\end{align}
where we introduce a dimensionless function $F(\nu, \bar a, \bar \lambda)$ with variables $\nu\equiv \omega/(\gamma_\eta a)$, $\bar a\equiv a\gamma_{\eta}^2$, and $\bar\lambda\equiv K\lambda/\gamma_{\eta}^3$.
Below, we show that the function $F(\nu,\bar a,\bar \lambda)$ behaves near the critical point as
\begin{align}
    F(\nu, \bar a, \bar\lambda) \xrightarrow[\bar a\to 0]{} \frac{4\pi}{3\bar a^{1/2}}\Phi(\nu),
\end{align}
independent of $\bar\lambda$ because of the critical slowing down of the soft mode $\psi$.

Using dimensionless quantities $x=p/a^{1/2}=p\xi$ and $\nu=\omega/(\gamma_{\eta} a) = \omega\xi^2/\gamma_{\eta}$, the first term in Eq.~\eqref{eq:loop_detail_photon} becomes
\begin{align}
\frac{T^2\gamma_{\eta}}{2\pi^2 K w} \int_0^{\infty}\frac{a^{3/2}x^2 dx}{a+ax^2} \int_{-1}^1 dz
\frac{2(\gamma_{\eta} a \nu)^2 - 4(\gamma_{\eta} a \nu) a^{1/2}x z + ax^2 + ax^2z^2}
{(\gamma_{\eta} a \nu)^2 + \left[
K\lambda ax^2 (a+ax^2) +
\gamma_{\eta} \{(\gamma_{\eta} a \nu)^2-2(\gamma_{\eta} a \nu) a^{1/2}xz + ax^2\}
\right]^2}.
\end{align}
Near the critical point $a\to 0$ while keeping $\nu\sim 1$, the integral is dominated by $x\sim 1$ region so that it is approximated by
\begin{align}
& \frac{T^2\gamma_{\eta}}{2\pi^2 K w a^{1/2}} \int_0^{\infty}\frac{x^2 dx}{1+x^2} \int_{-1}^1 dz
\frac{x^2 + x^2z^2}{\gamma_{\eta}^2 \nu^2 + \gamma_{\eta}^2 x^4} 
= \frac{4T^2}{3\pi^2 K w\gamma_{\eta}a^{1/2}} \int_0^{\infty}\frac{x^4 dx}{1+x^2}\frac{1}{\nu^2 + x^4}\propto a^{-1/2}.
\label{eq:loop_detail_photon_1st}
\end{align}
Similarly, the second term in Eq.~\eqref{eq:loop_detail_photon} is
\begin{align}
&\frac{T^2\lambda}{2\pi^2 w} \int_0^{\infty}dx \int_{-1}^1 dz
\frac{a^{5/2}x^4}{(\gamma_{\eta} a \nu)^2-2(\gamma_{\eta} a \nu) a^{1/2} xz + ax^2} \nonumber \\
&\qquad\qquad\times \frac{2(\gamma_{\eta} a \nu)^2 - 4(\gamma_{\eta} a \nu) a^{1/2}x z + ax^2 + ax^2z^2}
{(\gamma_{\eta} a \nu)^2 + \left[
K\lambda ax^2 (a+ax^2) +
\gamma_{\eta} \{(\gamma_{\eta} a \nu)^2-2(\gamma_{\eta} a \nu) a^{1/2}xz + ax^2\}
\right]^2}.
\end{align}
This integral is more complicated to analyze than the first one.
One needs to understand that the dominant contribution is from $x\sim (\gamma_{\eta}/K\lambda)^{1/2}a^{-1/2}$ instead of $x\sim 1$.
Then, by taking the limit $a\to 0$ while keeping $\nu\sim 1$, the integral is approximated by
\begin{align}
\frac{T^2\lambda a^{1/2}}{2\pi^2 w}
\int_0^{\infty}x^2 dx \int_{-1}^1 dz
\frac{x^2 + x^2z^2}
{\left(K\lambda ax^4 + \gamma_{\eta} x^2 \right)^2}
=\frac{4T^2\lambda}{3\pi^2 w}\int_0^{\infty}
\frac{a^{1/2}dx}{\left(K\lambda ax^2 + \gamma_{\eta} \right)^2}
\propto a^{0}.
\label{eq:loop_detail_photon_2nd}
\end{align}

Therefore, the first term ($\propto a^{-1/2}$) dominates the second term ($\propto a^0$) and we obtain
\begin{align}
\left(\delta_{ij}-\frac{k_ik_j}{\bm k^2}\right)\int\frac{d^4p}{(2\pi)^4} G_{S0}^{\psi\psi}(p)G_{S0}^{T ij}(k-p)
%\text{Eq.~\eqref{eq:loop}} 
&\simeq \frac{4T^2}{3\pi^2 K\bar\eta a^{1/2}} \int_0^{\infty}\frac{x^4 dx}{1+x^2}\frac{1}{\nu^2 + x^4}.
\end{align}
The last integration is calculated as follows
\begin{align}
\int_0^{\infty}\frac{x^4 dx}{1+x^2}\frac{1}{\nu^2 + x^4}
&=\frac{1}{2}\int_{-\infty}^{\infty}\frac{x^4 dx}{(x^2+1)(x^2-i\nu)(x^2+i\nu)}\nonumber\\
&=\frac{2\pi i}{2} \times\left[
\frac{i^4}{2i (-1-i\nu)(-1+i\nu)} 
+ \frac{(e^{i\pi/4}\sqrt{\nu})^4}{2e^{i\pi/4}\sqrt{\nu}(i\nu+1)(2i\nu)}
+ \frac{(e^{3i\pi/4}\sqrt{\nu})^4}{2e^{3i\pi/4}\sqrt{\nu}(-i\nu+1)(-2i\nu)}
\right]\nonumber \\
&=\frac{\pi}{2}\times
\frac{1-(1-\nu)\sqrt{\nu/2}}{1+\nu^2}
=\frac{\pi}{2}\Phi(\nu),
\end{align}
with which we finally obtain
\begin{align}
\text{\eqref{eq:PiT_oneloop}}
=\frac{{\mathcal A}^2}{4K}
\left(\delta_{ij}-\frac{k_ik_j}{\bm k^2}\right)\int\frac{d^4p}{(2\pi)^4} 
G_{S0}^{\psi\psi}(p)G_{S0}^{T ij}(k-p)
\simeq \frac{{\mathcal A}^2 T^2}{6\pi K\bar\eta a^{1/2}}\Phi(\nu).
\label{eq:scaling_damp_regime}
\end{align}
This derivation assumes a regime $\nu\sim 1$, i.e. $\omega\sim \gamma_{\eta}/\xi^2$.
It can be extended to a regime $\omega\sim 1/\xi$.
In this regime, the first term in the integral \eqref{eq:loop_detail_photon} is dominated by $x\sim \gamma_{\eta}^{-1/2}a^{-1/4}\gg 1$ instead of $x\sim 1$ while the second term is by the same range $x\sim (\gamma_{\eta}/K\lambda)^{1/2}a^{-1/2}$.
This difference only affects the approximate form of the integral \eqref{eq:loop_detail_photon_1st} by $x^2/(1+x^2)\simeq 1$ and thus the final formula \eqref{eq:scaling_damp_regime} is actually applicable to $\omega\sim 1/\xi$.

\section{Sound mode contributions to the photon production rate}
\label{app:sound_detail}
Expanding the baryon current up to the first nonlinear or dissipative corrections, we have as in Eq.~\eqref{eq:current_i}
\begin{align}
\bm J_B &\simeq \left[n + \left(\frac{\partial n}{\partial \hat s}\right)_{p}\delta\hat s
+\left(\frac{\partial n}{\partial p}\right)_{\hat s}\delta p\right] \bm v
-\sigma_B T \bm\nabla \left(\frac{\mu}{T}\right)  + \cdots.
\end{align}
Due to the transverse projection in the photon production rate, the sound mode contributes only from the one-loop order.
The nonlinear coupling between the sound mode $\bm v_L$ (longitudinal projection of $\bm v$) and the critical mode $\delta\hat s$ may enhance the photon production rate via the thermal loop effect.
Two-point function of momentum density is
\begin{align}
    G^{ij}_{S0}(p)&=
    \frac{2Tw^{-1}\gamma_\eta (\bm p^2\delta_{ij}-p_ip_j)}{(p^0)^2+\gamma_\eta^2 \bm p^4}
    +\frac{2Tw^{-1}\gamma_\zeta (p^0)^2 p_ip_j}{[(p^0)^2-c_s^2\bm p^2]^2+(\gamma_\zeta p^0\bm p^2)^2}
    = G^{Tij}_{S0}(p) + G^{Lij}_{S0}(p),
\end{align}
where $G^{Tij}_{S0}$ is the transverse mode contribution and $G^{Lij}_{S0}$ is the sound mode contribution.
We focus on the regime $|\bm p|\ll c_s/\gamma_{\zeta}$, where sound poles are well-defined peaks such that the correlator $G^{Lij}_{S0}$ can be split into
\begin{align}
    G^{Lij}_{S0}(p) 
    \simeq \frac{Tw^{-1}\gamma_{\zeta}p_ip_j}{(p^0-c_s|\bm p|)^2 + \gamma_{\zeta}^2 \bm p^4/4}
    +\frac{Tw^{-1}\gamma_{\zeta}p_ip_j}{(p^0+c_s|\bm p|)^2 + \gamma_{\zeta}^2\bm p^4/4}.
\end{align}
We calculate the one-loop integral
\begin{align}
&\left(\delta_{ij}-\frac{k_ik_j}{\bm k^2}\right)\int\frac{d^4p}{(2\pi)^4} G_{S0}^{\psi\psi}(p)G_{S0}^{Lij}(k-p) \nonumber \\
&=\left(\delta_{ij}-\frac{k_ik_j}{\bm k^2}\right)\int\frac{d^4p}{(2\pi)^4}
\frac{2T\lambda\bm p^2}{(p^0)^2 + (K\lambda)^2 \bm p^4(a + \bm p^2)^2}
\frac{Tw^{-1}\gamma_{\zeta}(k_i-p_i)(k_j-p_j)}
{(k^0-p^0 - c_s|\bm k-\bm p|)^2 + \gamma_{\zeta}^2 (\bm k-\bm p)^4/4} 
\quad + (c_s\to -c_s)\nonumber \\
&=\frac{2T^2\lambda\gamma_{\zeta}}{w}\int\frac{d^3 p}{(2\pi)^3}
\bm p^2 \left((\bm k-\bm p)^2 - \frac{(\bm k^2 - \bm k\cdot\bm p)^2}{\bm k^2}\right) \nonumber \\
&\qquad\times \left[
\frac{1}{2K\lambda \bm p^2(a + \bm p^2)} +\frac{1}{\gamma_{\zeta} (\bm k-\bm p)^2}
\right]
\frac{1}{(k^0 - c_s|\bm k-\bm p|)^2 + [K\lambda \bm p^2(a + \bm p^2) + \gamma_{\zeta} (\bm k-\bm p)^2/2]^2} \quad + (c_s\to -c_s) \nonumber \\
&=\frac{T^2\lambda\gamma_{\zeta}}{4\pi^2 w} \int_0^{\infty}p^2 dp \int_{-1}^1 dz
p^2 \left(k^2 - 2kpz + p^2 - (k-pz)^2\right)  
\left[
\frac{1}{K\lambda p^2(a + p^2)}+\frac{2}{\gamma_{\zeta} (k^2 -2kpz + p^2)}
\right]\nonumber \\
&\qquad\times\frac{1}{(k^0 - c_s\sqrt{k^2 -2kpz + p^2})^2 + [K\lambda p^2(a + p^2) + \gamma_{\zeta} (k^2 -2kpz + p^2)/2]^2} \quad  + (c_s\to -c_s).
\end{align}
From here, we use $\omega = k_0 = k$ for light-like external momentum:
\begin{align}
&\frac{T^2}{4\pi^2 K w} \int_0^{\infty} dp \int_{-1}^1 dz\left(
\frac{\gamma_{\zeta}p^2}{a + p^2}+\frac{2K\lambda p^4}{\omega^2 -2\omega pz + p^2}
\right)\nonumber \\
&\qquad\times\frac{p^2 \left(1 - z^2\right) }{(\omega - c_s\sqrt{\omega^2 -2\omega pz + p^2})^2
+ [K\lambda p^2(a + p^2) + \gamma_{\zeta} (\omega^2 -2\omega pz + p^2)/2]^2} \quad + (c_s \to -c_s).
\label{eq:loop_sound}
\end{align}
For $\omega \lesssim c_s^2/\gamma_{\zeta}$, the main contribution to this integral comes from $p\sim c_s /\gamma_{\zeta}\sim (\gamma_{\zeta}/K\lambda)^{1/2}$ and the result is $\propto a^0$ with no critical enhancement.
%The integral is numerically calculated in Fig.~\ref{fig:photon_sound} in the main text and compared with the model H results.
%We assume the transport coefficients satisfy $\bar\lambda\equiv K\lambda/\gamma_{\eta}^3=1$ and $\gamma_{\eta} = \gamma_{\zeta}/2$, and the sound velocity $c_s^2 = 1/3$.
%The overall magnitude is insensitive to $a$ and is typically much smaller than the model H results for $\omega\gamma_{\eta} \ll c_s^2$.

The integral \eqref{eq:loop_sound} does not diverge but has a physical cutoff scale at $p\sim c_s/\gamma_{\zeta}\sim (\gamma_{\zeta}/K\lambda)^{1/2}$ for hydrodynamic description.
We can estimate the integral for $\omega=0$ and $a=0$ by
\begin{align}
    &\frac{T^2}{4\pi^2 K w} \int_0^{c_s/\gamma_{\zeta}} dp \int_{-1}^1 dz
    \left(\gamma_{\zeta}+2K\lambda p^2\right)
    \frac{p^2 \left(1 - z^2\right) }{c_s^2 p^2 + (K\lambda p^4 + \gamma_{\zeta}p^2/2)^2} \times 2 \nonumber \\
    &\simeq \frac{T^2}{2\pi^2 K w}
    \int_0^{c_s/\gamma_{\zeta}} dp \int_{-1}^1 dz
    \left(\gamma_{\zeta}+2K\lambda p^2\right)\frac{\left(1 - z^2\right) }{c_s^2}
    =\frac{2T^2}{3\pi^2 K w} \left(
    \frac{1}{c_s} + \frac{2K\lambda c_s}{3\gamma_{\zeta}^3}
    \right)
    \simeq \frac{T^2}{K w} \times 0.12,
    \label{eq:sound_estimate}
\end{align}
where the final expression is obtained by substituting the values used in Fig.~\ref{fig:photon_sound}, i.e. $\bar\lambda\equiv K\lambda/\gamma_{\eta}^3=1$, $\gamma_{\eta}=\gamma_{\zeta}/2$, and $c_s^2 = 1/3$.
The main contribution to the integral \eqref{eq:loop_sound} comes from around the cutoff scale, resulting in a non-negligible deviation from our estimate.
To be specific, the estimate in \eqref{eq:sound_estimate} corresponds to the photon yield $\sim 10^{-5}$ at $\omega=0$, which is about 10\% of the numerical integration ($\sim 10^{-4}$) in Fig.~\ref{fig:photon_sound}.

\end{widetext}
\bibliography{Refs}

@article{Fukushima:2010bq,
    author = "Fukushima, Kenji and Hatsuda, Tetsuo",
    title = "{The phase diagram of dense QCD}",
    eprint = "1005.4814",
    archivePrefix = "arXiv",
    primaryClass = "hep-ph",
    reportNumber = "YITP-10-28, TKYNT-10-06",
    doi = "10.1088/0034-4885/74/1/014001",
    journal = "Rept. Prog. Phys.",
    volume = "74",
    pages = "014001",
    year = "2011"
}

@article{Bzdak:2019pkr,
    author = "Bzdak, Adam and Esumi, Shinichi and Koch, Volker and Liao, Jinfeng and Stephanov, Mikhail and Xu, Nu",
    title = "{Mapping the Phases of Quantum Chromodynamics with Beam Energy Scan}",
    eprint = "1906.00936",
    archivePrefix = "arXiv",
    primaryClass = "nucl-th",
    doi = "10.1016/j.physrep.2020.01.005",
    journal = "Phys. Rept.",
    volume = "853",
    pages = "1--87",
    year = "2020"
}

@article{STAR:2010vob,
    author = "Aggarwal, M. M. and others",
    collaboration = "STAR",
    title = "{An Experimental Exploration of the QCD Phase Diagram: The Search for the Critical Point and the Onset of De-confinement}",
    eprint = "1007.2613",
    archivePrefix = "arXiv",
    primaryClass = "nucl-ex",
    month = "7",
    year = "2010"
}

@article{Asakawa:1989bq,
    author = "Asakawa, M. and Yazaki, K.",
    title = "{Chiral Restoration at Finite Density and Temperature}",
    doi = "10.1016/0375-9474(89)90002-X",
    journal = "Nucl. Phys. A",
    volume = "504",
    pages = "668--684",
    year = "1989"
}

@article{Stephanov:2004wx,
    author = "Stephanov, Mikhail A.",
    editor = "Muller, Berndt and Tan, C. I.",
    title = "{QCD Phase Diagram and the Critical Point}",
    eprint = "hep-ph/0402115",
    archivePrefix = "arXiv",
    doi = "10.1143/PTPS.153.139",
    journal = "Prog. Theor. Phys. Suppl.",
    volume = "153",
    pages = "139--156",
    year = "2004"
}

@article{Stephanov:1998dy,
    author = "Stephanov, Misha A. and Rajagopal, K. and Shuryak, Edward V.",
    title = "{Signatures of the tricritical point in QCD}",
    eprint = "hep-ph/9806219",
    archivePrefix = "arXiv",
    reportNumber = "ITP-SB-98-39, MIT-CTP-2748, SUNY-NTG-98-17",
    doi = "10.1103/PhysRevLett.81.4816",
    journal = "Phys. Rev. Lett.",
    volume = "81",
    pages = "4816--4819",
    year = "1998"
}

@article{Stephanov:1999zu,
    author = "Stephanov, Misha A. and Rajagopal, K. and Shuryak, Edward V.",
    title = "{Event-by-event fluctuations in heavy ion collisions and the QCD critical point}",
    eprint = "hep-ph/9903292",
    archivePrefix = "arXiv",
    reportNumber = "ITP-SB-99-4, MIT-CTP-2834, SUNY-NTG-99-3",
    doi = "10.1103/PhysRevD.60.114028",
    journal = "Phys. Rev. D",
    volume = "60",
    pages = "114028",
    year = "1999"
}

@article{Berdnikov:1999ph,
    author = "Berdnikov, Boris and Rajagopal, Krishna",
    title = "{Slowing out-of-equilibrium near the QCD critical point}",
    eprint = "hep-ph/9912274",
    archivePrefix = "arXiv",
    reportNumber = "MIT-CTP-2931",
    doi = "10.1103/PhysRevD.61.105017",
    journal = "Phys. Rev. D",
    volume = "61",
    pages = "105017",
    year = "2000"
}

@article{Stephanov:2008qz,
    author = "Stephanov, M. A.",
    title = "{Non-Gaussian fluctuations near the QCD critical point}",
    eprint = "0809.3450",
    archivePrefix = "arXiv",
    primaryClass = "hep-ph",
    doi = "10.1103/PhysRevLett.102.032301",
    journal = "Phys. Rev. Lett.",
    volume = "102",
    pages = "032301",
    year = "2009"
}

@article{Asakawa:2019kek,
    author = {Asakawa, Masayuki and Kitazawa, Masakiyo and M\"uller, Berndt},
    title = "{Issues with the search for critical point in QCD with relativistic heavy ion collisions}",
    eprint = "1912.05840",
    archivePrefix = "arXiv",
    primaryClass = "nucl-th",
    reportNumber = "J-PARC-TH-0208",
    doi = "10.1103/PhysRevC.101.034913",
    journal = "Phys. Rev. C",
    volume = "101",
    number = "3",
    pages = "034913",
    year = "2020"
}

@article{RevModPhys.49.435,
  title = {Theory of dynamic critical phenomena},
  author = {Hohenberg, P. C. and Halperin, B. I.},
  journal = {Rev. Mod. Phys.},
  volume = {49},
  issue = {3},
  pages = {435--479},
  numpages = {0},
  year = {1977},
  publisher = {American Physical Society},
  doi = {10.1103/RevModPhys.49.435}
}

@article{PhysRevA.8.2586,
  title = {Dynamics of Fluids near the Critical Point: Decay Rate of Order-Parameter Fluctuations},
  author = {Swinney, Harry L. and Henry, Donald L.},
  journal = {Phys. Rev. A},
  volume = {8},
  issue = {5},
  pages = {2586--2617},
  numpages = {0},
  year = {1973},
  publisher = {American Physical Society},
  doi = {10.1103/PhysRevA.8.2586}
}

@article{KAWASAKI19701,
title = {Kinetic equations and time correlation functions of critical fluctuations},
journal = {Annals of Physics},
volume = {61},
number = {1},
pages = {1-56},
year = {1970},
issn = {0003-4916},
doi = {https://doi.org/10.1016/0003-4916(70)90375-1},
author = {Kyozi Kawasaki}
}

@article{PhysRevB.13.2110,
  title = {Renormalization-group treatment of the critical dynamics of the binary-fluid and gas-liquid transitions},
  author = {Siggia, E. D. and Halperin, B. I. and Hohenberg, P. C.},
  journal = {Phys. Rev. B},
  volume = {13},
  issue = {5},
  pages = {2110--2123},
  numpages = {0},
  year = {1976},
  publisher = {American Physical Society},
  doi = {10.1103/PhysRevB.13.2110}
}

@article{10.1143/PTP.55.1384,
    author = {Ohta, Takao and Kawasaki, Kyozi},
    title = "{Mode Coupling Theory of Dynamic Critical Phenomena for Classical Liquids. I: Dynamic Critical Exponents}",
    journal = {Progress of Theoretical Physics},
    volume = {55},
    number = {5},
    pages = {1384-1395},
    year = {1976},
    issn = {0033-068X},
    doi = {10.1143/PTP.55.1384}
}

@article{Son:2004iv,
    author = "Son, D. T. and Stephanov, M. A.",
    title = "{Dynamic universality class of the QCD critical point}",
    eprint = "hep-ph/0401052",
    archivePrefix = "arXiv",
    doi = "10.1103/PhysRevD.70.056001",
    journal = "Phys. Rev. D",
    volume = "70",
    pages = "056001",
    year = "2004"
}

@article{Bluhm:2020mpc,
    author = "Bluhm, Marcus and others",
    title = "{Dynamics of critical fluctuations: Theory \textendash{} phenomenology \textendash{} heavy-ion collisions}",
    eprint = "2001.08831",
    archivePrefix = "arXiv",
    primaryClass = "nucl-th",
    doi = "10.1016/j.nuclphysa.2020.122016",
    journal = "Nucl. Phys. A",
    volume = "1003",
    pages = "122016",
    year = "2020"
}

@article{Berges:2009jz,
    author = "Berges, Jurgen and Schlichting, Soren and Sexty, Denes",
    title = "{Dynamic critical phenomena from spectral functions on the lattice}",
    eprint = "0912.3135",
    archivePrefix = "arXiv",
    primaryClass = "hep-lat",
    doi = "10.1016/j.nuclphysb.2010.02.007",
    journal = "Nucl. Phys. B",
    volume = "832",
    pages = "228--240",
    year = "2010"
}

@article{Schlichting:2019tbr,
    author = {Schlichting, S\"oren and Smith, Dominik and von Smekal, Lorenz},
    title = "{Spectral functions and critical dynamics of the O(4) model from classical-statistical lattice simulations}",
    eprint = "1908.00912",
    archivePrefix = "arXiv",
    primaryClass = "hep-lat",
    doi = "10.1016/j.nuclphysb.2019.114868",
    journal = "Nucl. Phys. B",
    volume = "950",
    pages = "114868",
    year = "2020"
}

@article{Schweitzer:2020noq,
    author = {Schweitzer, Dominik and Schlichting, S\"oren and von Smekal, Lorenz},
    title = "{Spectral functions and dynamic critical behavior of relativistic $Z_2$ theories}",
    eprint = "2007.03374",
    archivePrefix = "arXiv",
    primaryClass = "hep-lat",
    doi = "10.1016/j.nuclphysb.2020.115165",
    journal = "Nucl. Phys. B",
    volume = "960",
    pages = "115165",
    year = "2020"
}

@article{Schweitzer:2021iqk,
    author = {Schweitzer, Dominik and Schlichting, S\"oren and von Smekal, Lorenz},
    title = "{Critical dynamics of relativistic diffusion}",
    eprint = "2110.01696",
    archivePrefix = "arXiv",
    primaryClass = "hep-lat",
    doi = "10.1016/j.nuclphysb.2022.115944",
    journal = "Nucl. Phys. B",
    volume = "984",
    pages = "115944",
    year = "2022"
}

@article{Florio:2021jlx,
    author = "Florio, Adrien and Grossi, Eduardo and Soloviev, Alexander and Teaney, Derek",
    title = "{Dynamics of the $O(4)$ critical point in QCD}",
    eprint = "2111.03640",
    archivePrefix = "arXiv",
    primaryClass = "hep-lat",
    doi = "10.1103/PhysRevD.105.054512",
    journal = "Phys. Rev. D",
    volume = "105",
    number = "5",
    pages = "054512",
    year = "2022"
}

@article{Florio:2023kmy,
    author = "Florio, Adrien and Grossi, Eduardo and Teaney, Derek",
    title = "{Dynamics of the O(4) critical point in QCD: Critical pions and diffusion in model G}",
    eprint = "2306.06887",
    archivePrefix = "arXiv",
    primaryClass = "hep-lat",
    doi = "10.1103/PhysRevD.109.054037",
    journal = "Phys. Rev. D",
    volume = "109",
    number = "5",
    pages = "054037",
    year = "2024"
}

@article{Nahrgang:2018afz,
    author = "Nahrgang, Marlene and Bluhm, Marcus and Schaefer, Thomas and Bass, Steffen A.",
    title = "{Diffusive dynamics of critical fluctuations near the QCD critical point}",
    eprint = "1804.05728",
    archivePrefix = "arXiv",
    primaryClass = "nucl-th",
    doi = "10.1103/PhysRevD.99.116015",
    journal = "Phys. Rev. D",
    volume = "99",
    number = "11",
    pages = "116015",
    year = "2019"
}

@article{Schaefer:2022bfm,
    author = "Schaefer, Thomas and Skokov, Vladimir",
    title = "{Dynamics of non-Gaussian fluctuations in model A}",
    eprint = "2204.02433",
    archivePrefix = "arXiv",
    primaryClass = "nucl-th",
    doi = "10.1103/PhysRevD.106.014006",
    journal = "Phys. Rev. D",
    volume = "106",
    number = "1",
    pages = "014006",
    year = "2022"
}

@article{Chattopadhyay:2023jfm,
    author = "Chattopadhyay, Chandrodoy and Ott, Josh and Schaefer, Thomas and Skokov, Vladimir",
    title = "{Dynamic scaling of order parameter fluctuations in model B}",
    eprint = "2304.07279",
    archivePrefix = "arXiv",
    primaryClass = "nucl-th",
    doi = "10.1103/PhysRevD.108.074004",
    journal = "Phys. Rev. D",
    volume = "108",
    number = "7",
    pages = "074004",
    year = "2023"
}

@article{Chattopadhyay:2024jlh,
    author = "Chattopadhyay, Chandrodoy and Ott, Josh and Schaefer, Thomas and Skokov, Vladimir V.",
    title = "{Simulations of Stochastic Fluid Dynamics near a Critical Point in the Phase Diagram}",
    eprint = "2403.10608",
    archivePrefix = "arXiv",
    primaryClass = "nucl-th",
    doi = "10.1103/PhysRevLett.133.032301",
    journal = "Phys. Rev. Lett.",
    volume = "133",
    number = "3",
    pages = "032301",
    year = "2024"
}

@article{Chattopadhyay:2024bcv,
    author = "Chattopadhyay, Chandrodoy and Ott, Josh and Schaefer, Thomas and Skokov, Vladimir V.",
    title = "{Critical fluid dynamics in two and three dimensions}",
    eprint = "2411.15994",
    archivePrefix = "arXiv",
    primaryClass = "nucl-th",
    month = "11",
    year = "2024"
}

@article{Basar:2024qxd,
    author = {Ba\c{s}ar, G\"ok\c{c}e and Bhambure, Jay and Singh, Rajeev and Teaney, Derek},
    title = "{Stochastic relativistic advection diffusion equation~from the Metropolis algorithm}",
    eprint = "2403.04185",
    archivePrefix = "arXiv",
    primaryClass = "nucl-th",
    doi = "10.1103/PhysRevC.110.044903",
    journal = "Phys. Rev. C",
    volume = "110",
    number = "4",
    pages = "044903",
    year = "2024"
}

@article{Bhambure:2024axa,
    author = "Bhambure, Jay and Mazeliauskas, Aleksas and Paquet, Jean-Francois and Singh, Rajeev and Singh, Mayank and Teaney, Derek and Zhou, Fabian",
    title = "{Relativistic Viscous Hydrodynamics in the Density Frame: Numerical Tests and Comparisons}",
    eprint = "2412.10303",
    archivePrefix = "arXiv",
    primaryClass = "nucl-th",
    month = "12",
    year = "2024"
}

@article{Bhambure:2024gnf,
    author = "Bhambure, Jay and Singh, Rajeev and Teaney, Derek",
    title = "{Stochastic relativistic viscous hydrodynamics from the Metropolis algorithm}",
    eprint = "2412.10306",
    archivePrefix = "arXiv",
    primaryClass = "nucl-th",
    month = "12",
    year = "2024"
}

@article{Stephanov:2017ghc,
    author = "Stephanov, M. and Yin, Y.",
    title = "{Hydrodynamics with parametric slowing down and fluctuations near the critical point}",
    eprint = "1712.10305",
    archivePrefix = "arXiv",
    primaryClass = "nucl-th",
    reportNumber = "MIT-CTP-4969, MIT-CTP/4969",
    doi = "10.1103/PhysRevD.98.036006",
    journal = "Phys. Rev. D",
    volume = "98",
    number = "3",
    pages = "036006",
    year = "2018"
}

@article{An:2019csj,
    author = {An, Xin and Ba\c{s}ar, G\"ok\c{c}e and Stephanov, Mikhail and Yee, Ho-Ung},
    title = "{Fluctuation dynamics in a relativistic fluid with a critical point}",
    eprint = "1912.13456",
    archivePrefix = "arXiv",
    primaryClass = "hep-th",
    doi = "10.1103/PhysRevC.102.034901",
    journal = "Phys. Rev. C",
    volume = "102",
    number = "3",
    pages = "034901",
    year = "2020"
}

@article{An:2020vri,
    author = {An, Xin and Ba\c{s}ar, G\"ok\c{c}e and Stephanov, Mikhail and Yee, Ho-Ung},
    title = "{Evolution of Non-Gaussian Hydrodynamic Fluctuations}",
    eprint = "2009.10742",
    archivePrefix = "arXiv",
    primaryClass = "hep-th",
    doi = "10.1103/PhysRevLett.127.072301",
    journal = "Phys. Rev. Lett.",
    volume = "127",
    number = "7",
    pages = "072301",
    year = "2021"
}

@article{Stephanov:2024mdj,
    author = "Stephanov, Mikhail",
    title = "{QCD Critical Point and Hydrodynamic Fluctuations in Relativistic Fluids}",
    eprint = "2403.03255",
    archivePrefix = "arXiv",
    primaryClass = "nucl-th",
    doi = "10.5506/APhysPolB.55.5-A4",
    journal = "Acta Phys. Polon. B",
    volume = "55",
    number = "5",
    pages = "5--A4",
    year = "2024"
}

@article{Chandran:2012cjk,
    author = "Chandran, Anushya and Erez, Amir and Gubser, Steven S. and Sondhi, S. L.",
    title = "{Kibble-Zurek problem: Universality and the scaling limit}",
    doi = "10.1103/PhysRevB.86.064304",
    journal = "Phys. Rev. B",
    volume = "86",
    number = "6",
    pages = "064304",
    year = "2012"
}

@article{Mukherjee:2015swa,
    author = "Mukherjee, Swagato and Venugopalan, Raju and Yin, Yi",
    title = "{Real time evolution of non-Gaussian cumulants in the QCD critical regime}",
    eprint = "1506.00645",
    archivePrefix = "arXiv",
    primaryClass = "hep-ph",
    doi = "10.1103/PhysRevC.92.034912",
    journal = "Phys. Rev. C",
    volume = "92",
    number = "3",
    pages = "034912",
    year = "2015"
}

@article{Mukherjee:2016kyu,
    author = "Mukherjee, Swagato and Venugopalan, Raju and Yin, Yi",
    title = "{Universal off-equilibrium scaling of critical cumulants in the QCD phase diagram}",
    eprint = "1605.09341",
    archivePrefix = "arXiv",
    primaryClass = "hep-ph",
    doi = "10.1103/PhysRevLett.117.222301",
    journal = "Phys. Rev. Lett.",
    volume = "117",
    number = "22",
    pages = "222301",
    year = "2016"
}

@article{Akamatsu:2018vjr,
    author = "Akamatsu, Yukinao and Teaney, Derek and Yan, Fanglida and Yin, Yi",
    title = "{Transits of the QCD critical point}",
    eprint = "1811.05081",
    archivePrefix = "arXiv",
    primaryClass = "nucl-th",
    reportNumber = "MIT-CTP-5042, MIT-CTP/5042",
    doi = "10.1103/PhysRevC.100.044901",
    journal = "Phys. Rev. C",
    volume = "100",
    number = "4",
    pages = "044901",
    year = "2019"
}

@article{Nishimura:2023oqn,
    author = "Nishimura, Toru and Kitazawa, Masakiyo and Kunihiro, Teiji",
    title = "{Enhancement of dilepton production rate and electric conductivity around the QCD critical point}",
    eprint = "2302.03191",
    archivePrefix = "arXiv",
    primaryClass = "hep-ph",
    reportNumber = "YITP-23-12, J-PARC-TH-0283",
    doi = "10.1093/ptep/ptad051",
    journal = "PTEP",
    volume = "2023",
    number = "5",
    pages = "053D01",
    year = "2023"
}

@article{Nishimura:2024kvz,
    author = "Nishimura, Toru and Kitazawa, Masakiyo and Kunihiro, Teiji",
    title = "{Electromagnetic response of dense quark matter around color-superconducting phase transition and QCD critical point}",
    eprint = "2405.09240",
    archivePrefix = "arXiv",
    primaryClass = "hep-ph",
    reportNumber = "YITP-24-63, J-PARC-TH-0305",
    doi = "10.1016/j.aop.2024.169768",
    journal = "Annals Phys.",
    volume = "469",
    pages = "169768",
    year = "2024"
}

@article{Arnold:2001ms,
    author = "Arnold, Peter Brockway and Moore, Guy D. and Yaffe, Laurence G.",
    title = "{Photon emission from quark gluon plasma: Complete leading order results}",
    eprint = "hep-ph/0111107",
    archivePrefix = "arXiv",
    reportNumber = "UW-PT-01-22",
    doi = "10.1088/1126-6708/2001/12/009",
    journal = "JHEP",
    volume = "12",
    pages = "009",
    year = "2001"
}

@book{LeBellac_1996, 
place={Cambridge}, 
series={Cambridge Monographs on Mathematical Physics}, 
title={Thermal Field Theory}, 
publisher={Cambridge University Press}, 
author={Le Bellac, Michel}, year={1996}, 
collection={Cambridge Monographs on Mathematical Physics}
}

\end{document}